\documentclass[vecphys,pdftex]{svmult}

\usepackage{graphicx}        
\usepackage[bottom]{footmisc}

\usepackage{ulem}            

\def\msun{M$_{\odot}$}
\def\rsun{R$_{\odot}$}

\def\sbs{SBS\,1150+599A}
\def\pn{PN\,G135.9+55.9}

\def\it{\sl}
\def\degs{\ifmmode ^{\circ}\else$^{\circ}$\fi}
\def\amin{\ifmmode ^{\prime}\else$^{\prime}$\fi}
\def\asec{\ifmmode ^{\prime\prime}\else$^{\prime\prime}$\fi}

\def\degs{\ifmmode ^{\circ}\else$^{\circ}$\fi}
\def\amin{\ifmmode ^{\prime}\else$^{\prime}$\fi}

\def\eqalign#1{\null\,\vcenter{\openup1\jot \m@th
   \ialign{\strut\hfil$\displaystyle{##}$&$\displaystyle{{}##}$\hfil
   \crcr#1\crcr}}\,}
\begin{document}

\title*{An object that defies stereotypes.\\  {\small X-ray observations of SBS\,1150+599A -- the binary nucleus of PN\,G135.9+55.9}}
 \titlerunning{An object that defies stereotypes} 
\author{G.Tovmassian\inst{1},
J.Tomsick\inst{2}, R.Napiwotzki\inst{3},  L.Yungelson\inst{4},\\ G.Stasi\'nska\inst{5}, M.Pe\~na\inst{1} \and M.Richer\inst{1}}
\authorrunning{Tovmassian el al} 
\institute{Institute of Astronomy, Universidad Nacional Autonoma de M\'exico.
\texttt{gag,richer@astrosen.unam.mx, miriam@astroscu.unam.mx}
\and University California Berkeley, USA \texttt{jtomsick@ssl.berkeley.edu}
\and University of Hertfordshire, UK \texttt{r.napiwotzki@herts.ac.uk}
\and Institute of Astronomy, RAS, Russia \texttt{lry@inasan.ru}
\and Observatoire du Meudon, France \texttt{grazyna.stasinska@obspm.fr}}
%
%
\maketitle

\begin{abstract}
We present X-ray observations of the close binary nucleus of the planetary nebula (PN) PN\,G135.9+55.9 obtained with the XMM satellite. The nebula is  the most oxygen-poor PN known to date and is located in the Galactic halo. It is known to harbor a close binary nucleus of which only one component can be observed in optical-UV range. New X-ray observations show that the invisible component is a very hot compact star. This finding allows us to reconstruct the immediate past of the object and predict its future. The parameters of the binary components we determine strongly suggest that the precursor was a symbiotic supersoft X-ray source that finished its life by Roche lobe overflow. PN\,G135.9+55.9 is an excelent candidate for a future type Ia supernova.
\end{abstract}

\section{Introduction}
\label{sec:1}
\sbs\ was identified as a planetary nebula (PN) in \cite{2001A&A...370..456T} and subsequently designated as \pn. The object is unusual  and is renown for its extremely low oxygen content \cite{2001A&A...370..456T, 2002AJ....124.3340J, 2005AA...430..187P,2005IAUS..228..323S}. It is also located far and above the Galactic plane, which places it among  a handful of known halo PNe. Direct images obtained by \cite{2002AA...395..929R,2002AJ....124.3340J} and a study of dynamics of the nebula \cite{2003AA...410..911R} confirmed its PN identification, but did not make its explanation easier.   Another, if not unusual but outstanding feature of the PN is that it harbors a close binary system,  only one component of which is observable  in the optical and UV \cite{2004ApJ...616..485T}.  The far-UV observations by FUSE established a temperature range for that visible  component of 110\,000 - 120\,000\,K. The lower limit of this range was established by a minimum temperature  needed to produce the [Ne\,V] emission line, while the upper limit was deduced from a continuum fit of atmosphere models to the observed data \cite{2004ApJ...616..485T}. 
The extremely high temperature coupled with high $log$\,g 
led us to assume that the opt/UV component was   the central star of planetary nebula (CSPN), i.e. the post-AGB star that lost its envelope and was the source of its ionization.
P{\'e}quignot  \& Tsamis \cite{2005AA...430..187P} claimed that the ionizing star must be hotter in order to increase the oxygen abundance to  more common levels ([O/H]$\approx -0.9$ instead of [O/H]$\le -2.2$). It was an extreme solution:  to justify a temperature of  120\,000\,K  \cite{2004ApJ...616..485T}  a higher extinction correction  than  estimated for the direction of \pn\ is required.  
To resolve this controversy, we \cite{2005AIPC..804..173N} obtained photometric light curves of the binary core of \pn. The orbital period of the system is 3.919 hr, and to explain the double-humped shape of the light curve, one should invoke a Roche-lobe filling opt/UV component. Additionally, the depths of the minima in the light curve are uneven, which in turn assumes that the visible component should be irradiated by an even more energetic (hot)  source.  The dynamics required that this invisible component must be another compact object  of at least 0.85\msun\  \cite{2005AIPC..804..173N}.

Here we report the analysis and results of X-ray observations of \sbs, the close binary core of \pn\ made with orbital telescope XMM-{\sl Newton} and make use of publicly available HST UV-data. 

\section{Observations and spectral energy distribution (SED).}
\label{sec:2}
On 2006 Nov 1, a 25 ksec exposure of \sbs\ was obtained with  XMM-{\sl Newton}. We reduced the observations using the latest version of SAS7.0. The object was detected with XMM with a count rate of 0.035 c/s in the PN detector, 0.0029 c/s in the MOS1 detector, and 0.0048 c/s in the MOS 2 detector. \sbs\ appears to be emitting only very soft photons, the  absolute majority of which were registered in a narrow range between 0.1 to 0.3 keV. This soft end of spectral range suffers calibration problems \cite{2006ESASP.604..937S},  but the extracted spectra in all three detectors  are highly consistent with each other and are well-described with a black-body (BB). 

For a better understanding of the nature of the object and comprehensive fit of the SED we also use the UV data obtained by HST in 2003 (Obs. ID \#9466, PI Garnavich). Here, we use the continuum fit to the stellar spectra of the object obtained by STIS in near and far UV ranges. The STIS spectrum helps to fill the gap between optical and extremely far UV data obtained previously with FUSE. More importantly it eliminates the large uncertainty regarding extrapolation of the interstellar  extinction law in the FUSE range and in general provides a more reliable flux calibration. Thus, we have observational data continuously covering the whole spectral range from about 10\,000 to 900\AA.

%
%
%
\begin{figure}[t]
\centering
\includegraphics*[scale=0.55,bb=50 200 600 600,clip=]{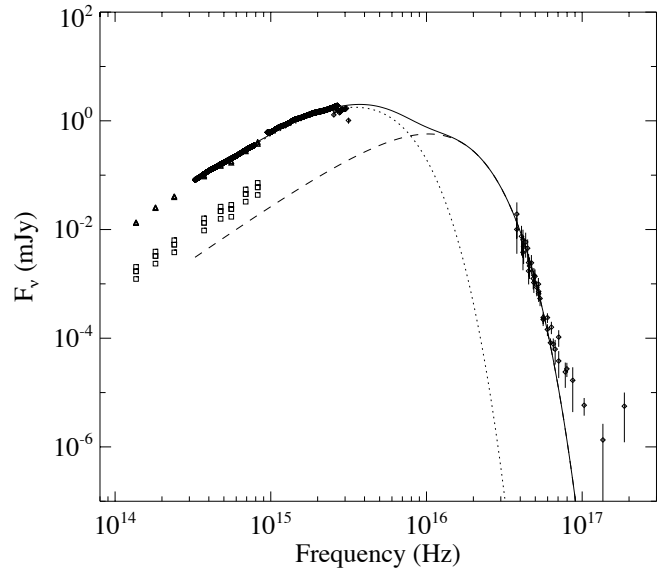}
%
%
\caption{Spectral energy distribution of \sbs. Open circles with error bars are X-ray data. Filled and open diamonds  are optical \& UV data. The observed data are fitted with two blackbodies presented with dotted \& dashed lines and their sum by a solid line.  The open triangles and open squares are SED of opt/UV and X-component respectively calculated by {\sl Nightfall} for a number of solutions. They all are adjusted to the flux of  opt/UV component. The discrepancy between {\sl Nightfall}  and double BB fits can be explained by lack of consideration of irradiation, geometry and effects of limb-darkening etc., in BB fitting.  The  solutions with much  hotter X-component produce a larger discrepancy and  are ruled out.}
\label{fig:1}       
\end{figure}

The SED  over a wide range of frequencies from infrared (IR) to X-rays  is presented in Fig~\ref{fig:1}. The optical and UV data can be fitted with a single black body (BB), while, for the X-rays, a second  BB with much higher temperature  is required. The discovery of a second, hotter component in the system means that we do not need to artificially elevate the extinction towards  \sbs\ in order to justify the presence of an ionization source of at least 110\,000\,K. Therefore in a revision of previous results reported in \cite{2004ApJ...616..485T} we apply a regular extinction in this direction  E(B-V)=0.029 \cite{1998ApJ...500..525S}  and fit the observed points with two BB using a $\chi^2$ method. One of the possible solutions is presented in Fig.~\ref{fig:1}. Other solutions with comparable $\chi^2$ are viable by increase of the temperature/decrease of size of the X-ray component, due to some excess of emission  at the  high-energy  end of the spectrum. We explain below why we chose 
the solution with the lower temperature. 
According to this fit the opt/UV component has a temperature of $\approx58\,000$\,K and ratio of the radius to the distance (r/D)$_{\rm opt/UV}=3.8\times10^{-13}$, while the X-ray component has T$_{X}=170\,000$\,K and (r/D)$_{\rm X}=5.6\times10^{-14}$. 

\section{Parameters and the nature of the system.}
\label{sec:3}

As the next step, we may try to determine binary system parameters, making use of the radial velocity (RV) curve and lightcurves in different colors. The rough knowledge of  component's  temperatures
greatly improves our chances of finding other parameters of the system correctly. For this purpose we use a binary modeling code {\it Nightfall} provided by R. Wichmann.\footnote{ http://www.hs.uni-hamburg.de/DE/Ins/Per/Wichmann/Nightfall.html.\\
{\it Nightfall} is based on a physical model that takes into account the nonspherical shape of stars in close binary systems, as well as mutual irradiance of both stars, and a number of additional physical effects, like albedo or limb darkening. {\sl Nightfall} can handle a large range of binary star configurations, including overcontact (common envelope) systems, eccentric orbits, surface spots and 
synchronous rotation, and the possible existence of a third star in the system.} The program allows fitting a large number of unknown parameters of the system: total mass, mass ratio of components, size of the components measured in  fractions  of corresponding Roche lobe, temperatures and separation of the system components. As an input for fitting we provided RV curve as measured in \cite{2004ApJ...616..485T} and light curves in three colors (roughly corresponding to {\it U, B, R} described in \cite{2005AIPC..804..173N}). Setting all parameters free forces the program immediately to seek solutions in a high mass stellar binary ($>20$\msun). However we know for sure that the system is comprised  of compact objects (WD-like) so the total mass should hover
around 1.4\msun. Fixing the total mass and temperature(s) of the components narrows the number of possible solutions, but a variety of different configurations result in a similarly good fit to the data. Therefore,  additional constraints are necessary to pin down the correct composition of \sbs.

 %
%
%
\begin{figure}[t]
\centering
\includegraphics[height=5.cm]{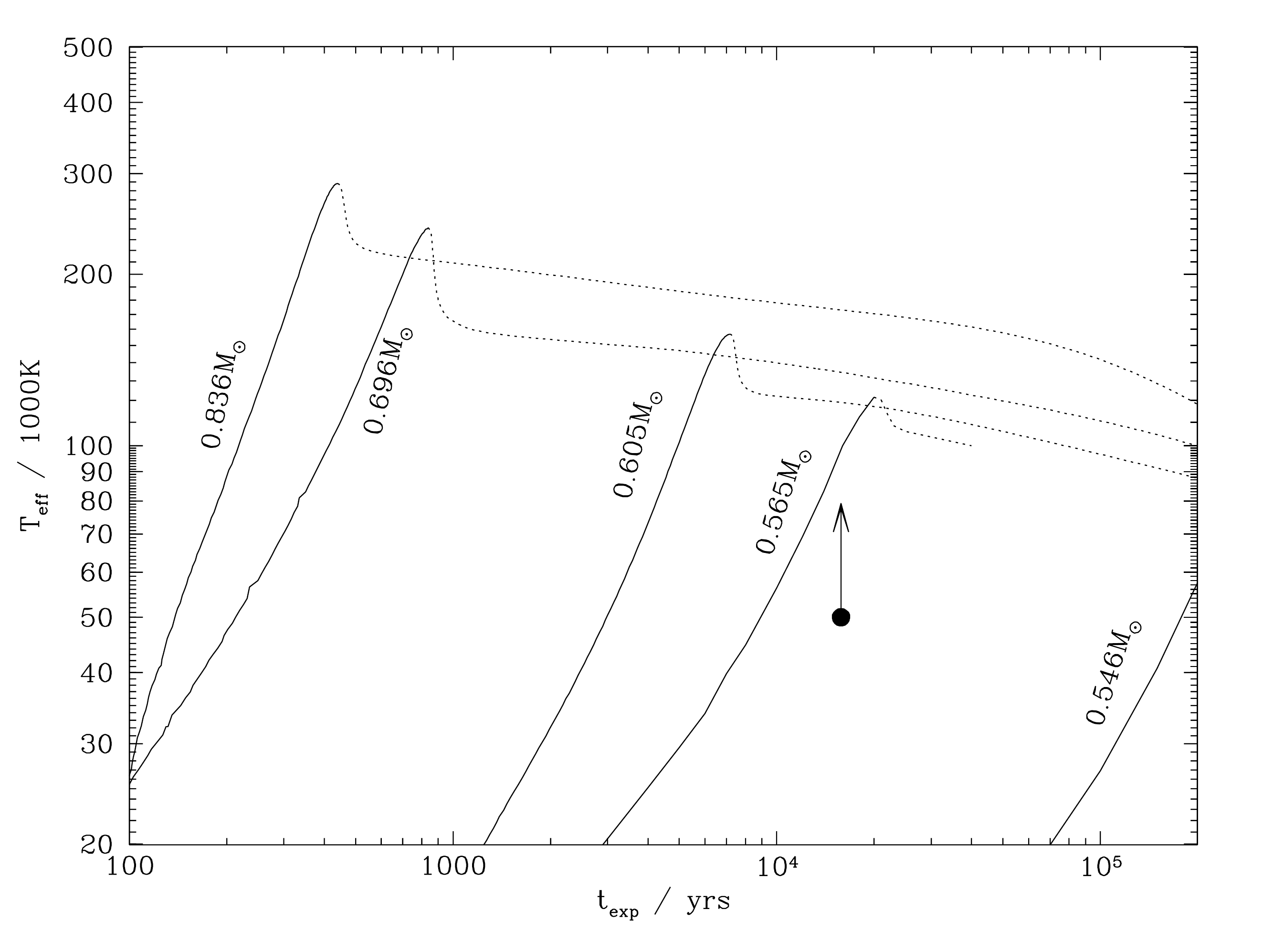}
%
%
\caption{Evolution tracks of low mass post-AGB stars according to \cite{1995A&A...299..755B}. The place of opt/UV component of \sbs\ is marked by an arrow.}
\label{fig:2}       
\end{figure}

\begin{figure}[t]
\centering
%
\includegraphics[height=4.5cm]{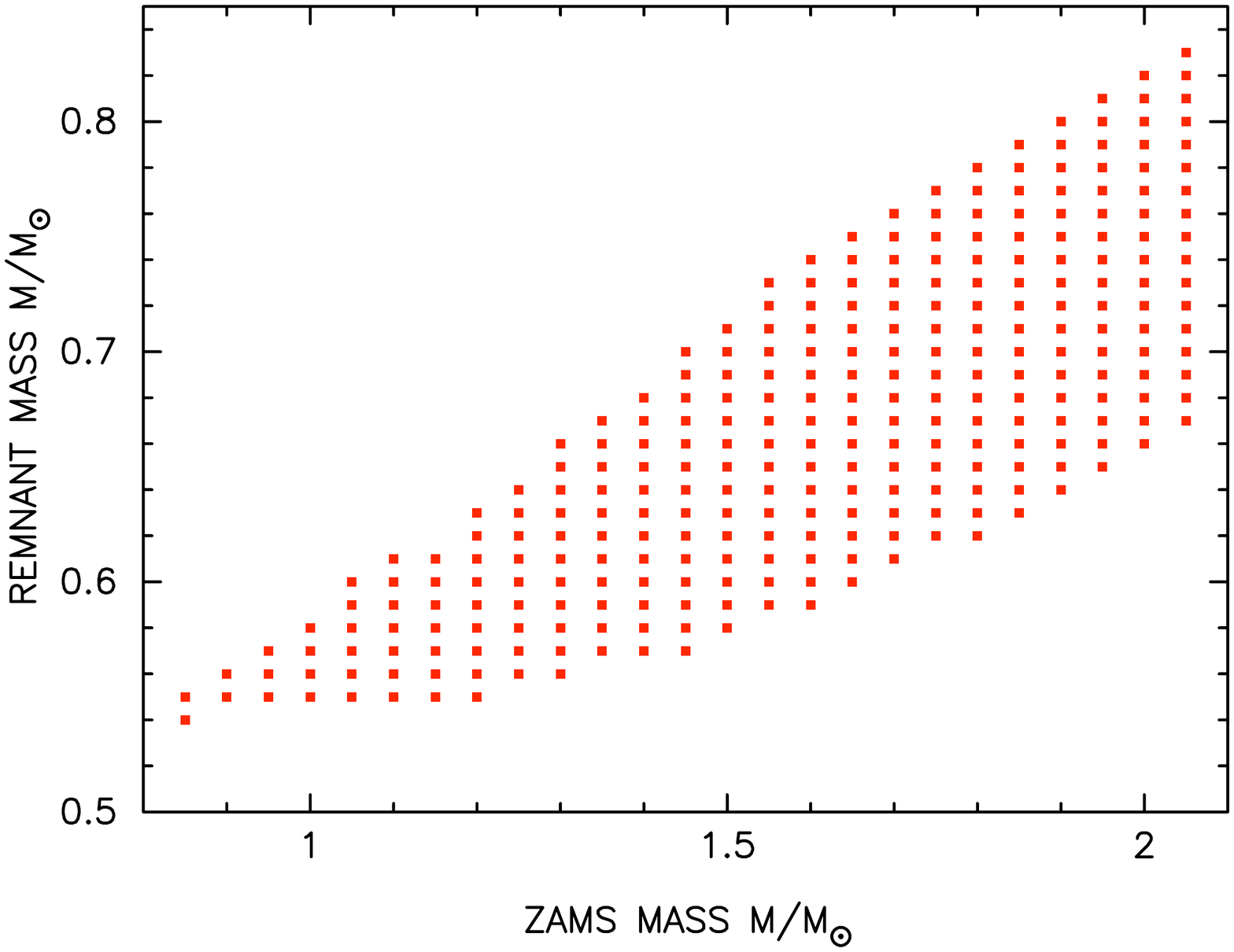}
\includegraphics[height=4.5cm]{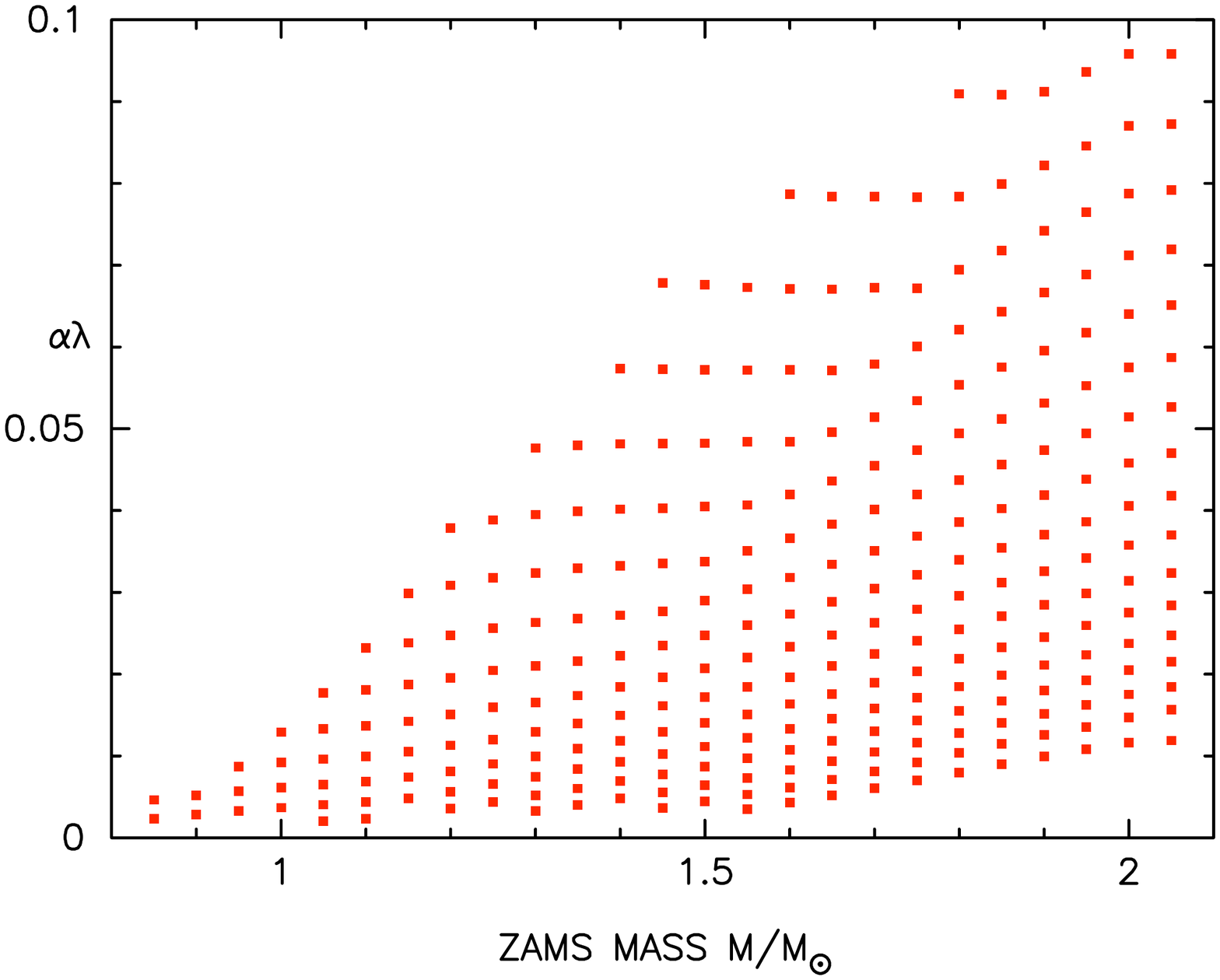}
%
%
\caption{Possible combinations of initial and final masses of ${\rm opt/UV}$ component
of \sbs\ (left panel) and coresponding $\alpha\lambda$ combinations of binding energy $\lambda$ \& expulsion efficiency $\alpha_{\rm CE}$ that result in orbital period of 3.9 hr (right panel).}   
\label{fig:3}       
\end{figure}

 One of the most important parameters to know are the masses of the components. We derive the mass of the opt/UV component from the fact that it is the  envelope-shedding component on its track from AGB to WD. 
 According to \cite{1995A&A...299..755B}, a core of post-AGB  star can reach temperatures $\ge55\,000$\,K in a reasonable time ($<  \rm {a\, few} \times10^4$ yr)  only  if it mass exceeds 0.56\msun\  (Fig~\ref{fig:2}). On the other hand we can evaluate the upper limit to the mass of opt/UV component from the consideration of common envelope (CE) evolution. In an 3.9\,hr Pop. II binary,  a 0.56\msun\  visible component may descend from AGB star that had ZAMS  mass 0.9-1.1\,\msun. Formation of a double degenerate  with given orbital period
 through common envelope requires a certain  combination of the binding energy $\lambda$ and  CE expulsion efficiency $\alpha_{\rm CE}$, e.g. 
 \cite{2007arXiv0704.0280W}.     
 We searched solutions for a range of progenitor masses  using code \cite{2000MNRAS.315..543H}
  like in \cite{2007A&A...466.1031V}. 
The grid of solutions is presented in Fig~\ref{fig:3}. 
The mass of the  ${\rm opt/UV}$ component progenitor should not  exceed 1.1\msun, provided that  \sbs\  is Pop. II object, and it is not a strange invader captured by merging of a sattelite galaxy. Fig~\ref{fig:3} shows that this results in  $M_{\rm opt/UV} < 0.58$\msun. These mass limits correspond to $\alpha_{\rm CE}\lambda= 0.01-0.02$, which is not unreasonable for stars with relatively large cores and tenuous envelopes. 
 
 %
%
%

We also can establish  the  lower limit  for the  mass of the X-ray component. Apparently the X-ray component is a WD remnant of  a star, which was more massive than the progenitor of opt/UV component and has evolved through its AGB phase much earlier. The system most probably passed through the first CE phase  and for a certain time appeared as a symbiotic system, see, e. g.,  \cite{2006MNRAS.372.1389L}. In  the symbiotic  phase, the X-component  could accrete mass  at a  high rate ($\sim 10^{-7}$\msun/yr) and manifest itself as a supersoft X-ray source (SSS), see, e. g.  \cite{2006AdSpR..38.2836K}.  The accreting WD in such systems gets heated to extremely high temperatures by  steady nuclear burning of accreted matter at their  surface. Only accretors with M$_{\rm wd}>0.7$\msun\  reach temperatures high enough to be detected as SSS \cite{2001ASPC..229..309W}. 
 The estimate of amount of matter accreted during symbiotic phase varies (\cite{2007AIPC..924..903K,2006MNRAS.372.1389L,2002ASPC..261..605H}), but at least 0.1-0.15\msun\ must be accumulated as a result.
Therefore, we can  assume that the  total mass of the system is very close to or exceeding the Chandrasekhar limit. 

Introducing these additional constraints to the {\sl Nightfall}, we find the best solutions  describing the light and radial velocity curves at the same time. The parameters of the system deduced by our analysis are presented  in the Table~\ref{tab:1}.  

%
%
\begin{table}
\centering
\caption{The parameters of binary components of \sbs.}
\label{tab:1}       
%
%
\begin{tabular}{lll}
\hline\noalign{\smallskip}
 & opt/UV-component & X-component  \\
\noalign{\smallskip}\hline\noalign{\smallskip}
M (\msun)\, & 0.565 & 0.85  (M$_{\rm tot} = 1.41$\msun) \\
R (\rsun) & 0.40 -- 0.48 & 0.16 -- 0.05 \\
T (K) &    65 - 57\,000 &  165 --  245\,000\\
\noalign{\smallskip}\hline
\end{tabular}
\end{table}
%
%
%
\begin{figure}[t]
\centering
\includegraphics[height=3.5cm,bb=200 200 700 500,clip=]{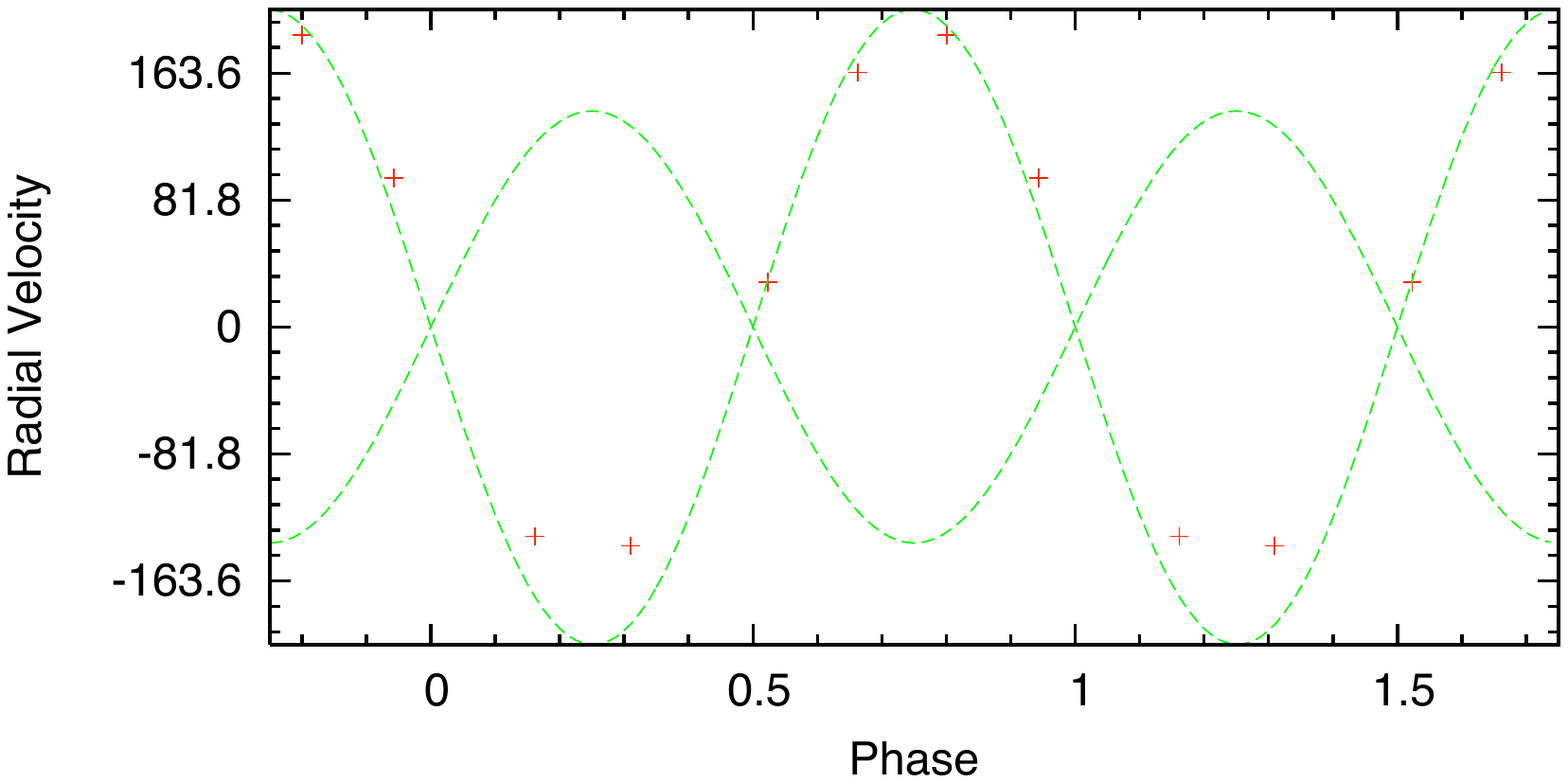}
\includegraphics[height=3.5cm,bb=200 200 700 500,clip=]{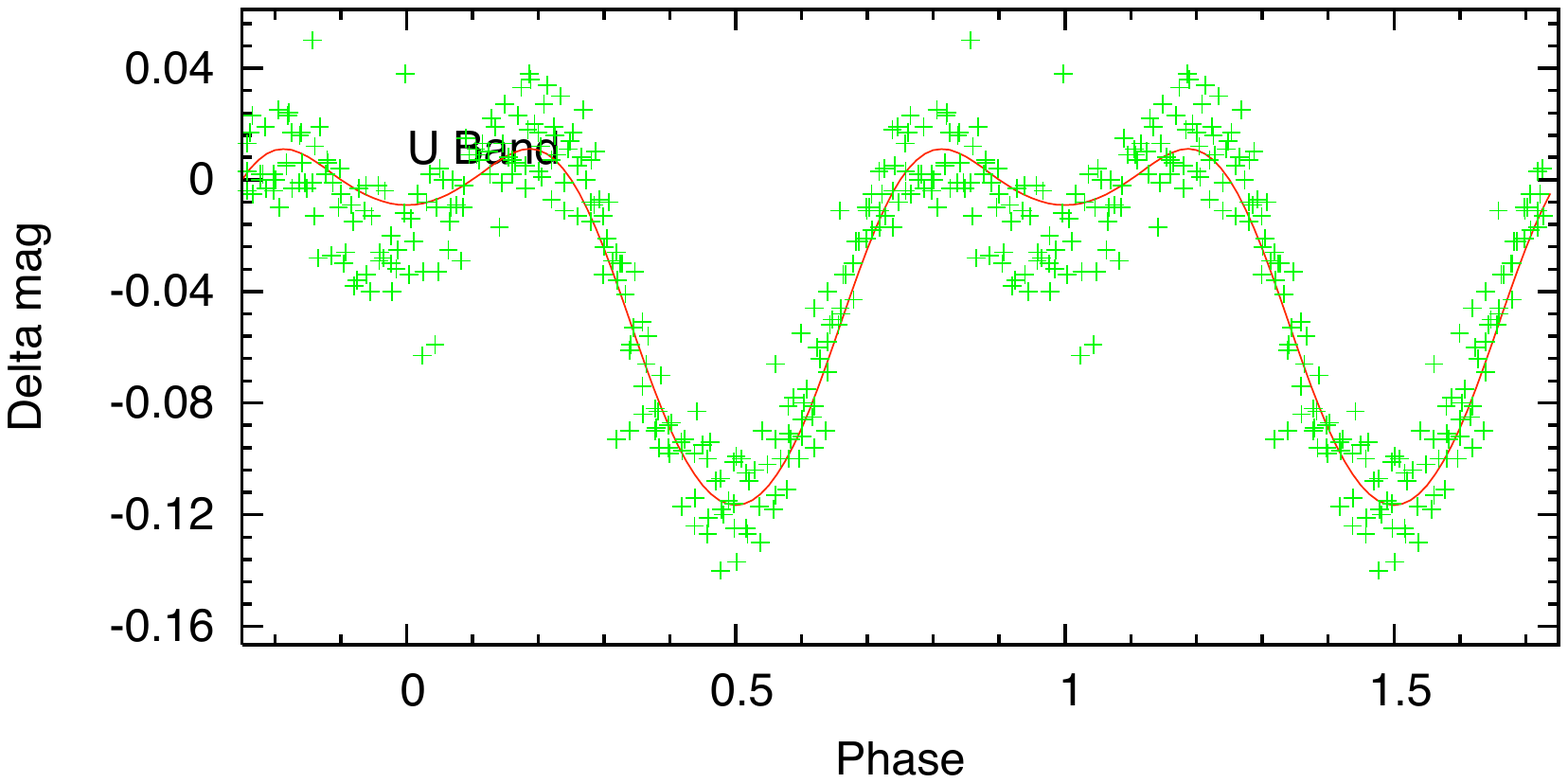}
\includegraphics[height=3.5cm,bb=200 200 700 500,clip=]{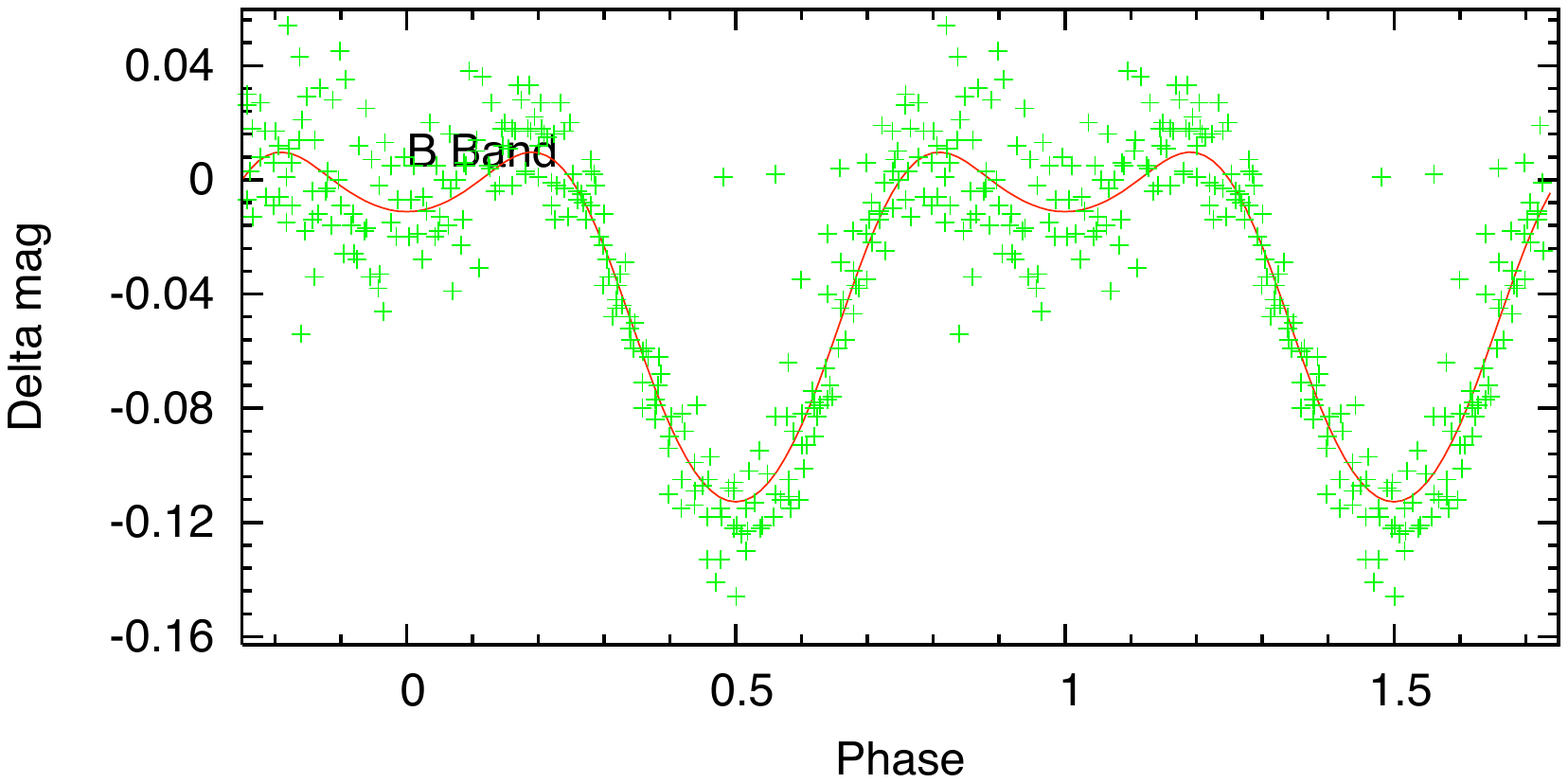}
\includegraphics[height=3.5cm,bb=200 200 700 500,clip=]{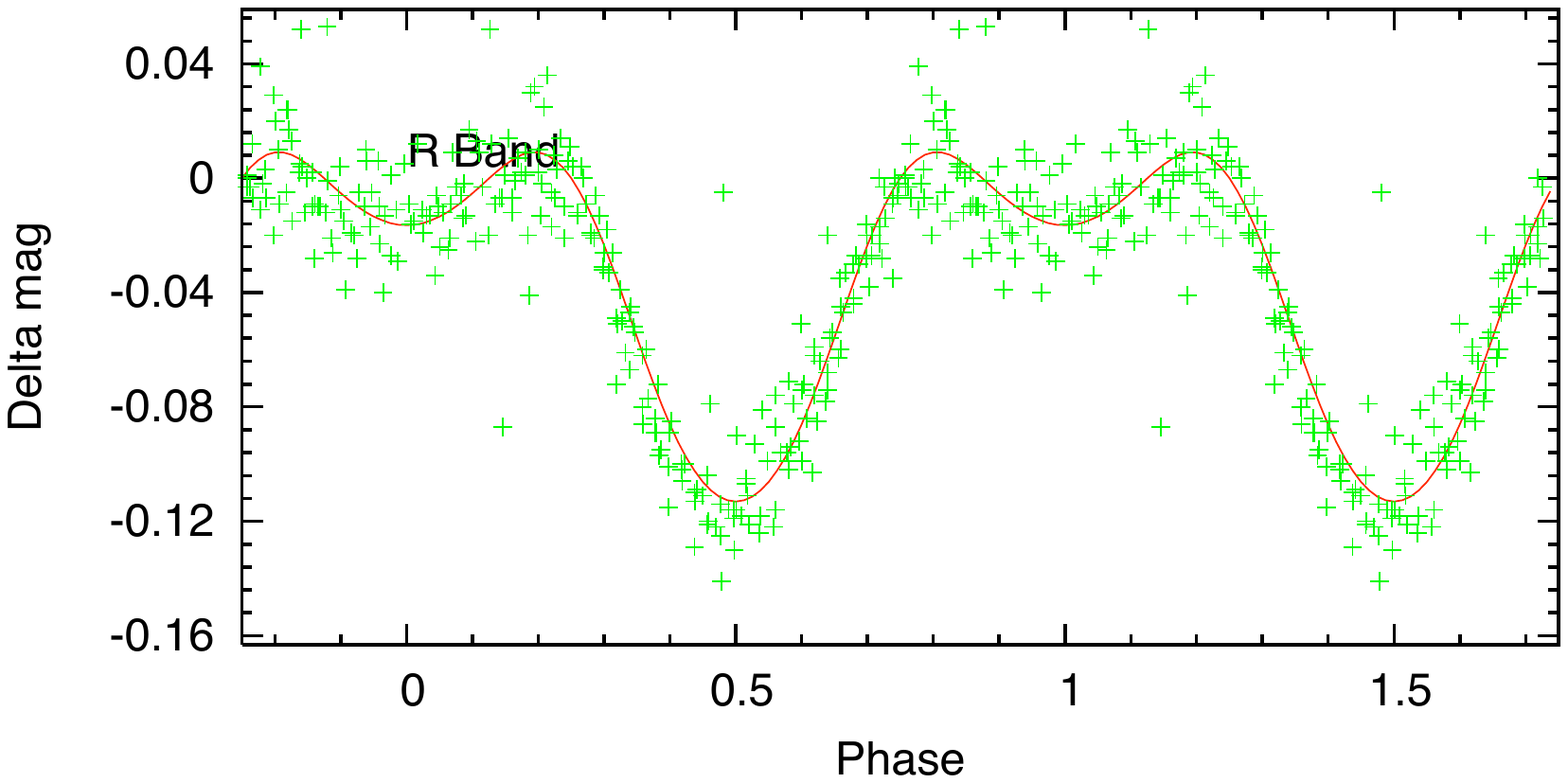}
%
%
\caption{Simultaneous fits to RV and light curves in three colors by {\sl Nightfall}. The parameters of the system deduced from $\chi^2$ fits are presented in Tab~\ref{tab:1}}
\label{fig:4}       
\end{figure}

Corresponding fits to the RV and light curves are presented in Fig~\ref{fig:4}. The best fits are achieved with inclination angle $i= 50\deg \pm4$. The wide range of values in the right column of Tab.~\ref{tab:1} indicates that similarly good fits can be obtained with a variety of temperature/radius combination for the X-component. What matters for the fits is the luminosity of  X-component  to ensure certain irradiation of the Roche-lobe filling opt/UV-component. Thus, the X-ray component may be cooler and larger or hotter and smaller. But, regardless of the uncertainties of the XMM calibration and the double blackbody fitting to the SED,  the correct solution is a rather cool and large X-ray component, because the flux ratio of the components should obey the ratio calculated by {\sl Nightfall} for the Raleigh-Jeans tail and presented in Fig~\ref{fig:1} by open triangles for the opt/UV and by the open squares for X-ray component. An extremely hot solution clearly deviates from the presented set of solutions. However, even a hot solution requires a size obviously larger than an ordinary WD. It confirms the  suggestion  that the X-component underwent a period of intense accretion and  is surrounded by a shell-like atmosphere, heated  by the surface burning.

\section{Conclusions.}

%
%
\begin{figure}[t]
\centering
\includegraphics[height=7.5cm]{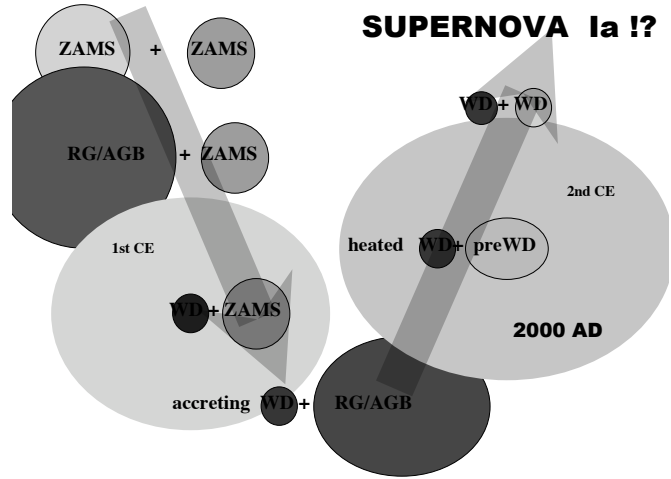}
%
%
\caption{Schematic evolutional trajectory of \sbs. The current state of the object is marked by  2000AD. }
\label{fig:5}       
\end{figure}

New X-ray observations coupled with the previous UV and optical spectral energy distribution reveal a second, previously undetected component, of the close double degenerate binary core of \pn.  It appears to be a WD that has grown in size from accreted material in a previous symbiotic/SSS stage and heated by a steady nuclear burning on its surface. The discovery of  the X-ray component adds another unprecedented feature to the already unique characteristics of the object, but at the same time helps to explain its nature. According to some scenarios,  double degenerate compact binaries, progenitors of  SNIa are formed as a result of two common envelope phases, which helps to remove significant amounts of angular momentum and bring the  remnants of the binary system  components close enough to merge in a Hubble time.  We suggest that in rare cases these  systems pass through a symbiotic stage and  show up for a short time as  supersoft X-ray sources.
 A schematic evolutionary scenario for such systems is presented  in Fig~\ref{fig:5}. Currently, we detect a PN formed as a CE released by opt/UV-component. But its ionization is due to the extreme UV photons from both components of the binary system, which means  that the oxygen abundance should be revised.
 \sbs\  remains, however, one of the rarest objects of its type, observed as a PN, heated  by its extremely hot binary core,  with an extreme chemical composition, located in the very outskirts of the Galaxy. Its observation over  a wide range of wavelengths  allows  insight  into one of the peculiar phases of evolution that lasts only a short time and thus is difficult to catch.
\smallskip 

\subparagraph{Acknowledgments} {\small GT \& JT acknowledge NASA grant NNX07AQ12G , GT is also funded by CONACyT. LY is supported by ``Origin and Evolution of Stars and 
Galaxies'' Program and RFBR grant 07-02-00454. We appreciate expertise and macros rendered by J.Wilms for XMM data reduction. 
}
%
%



\end{document}